\documentclass[preprint,pre]{revtex4}
\usepackage[pdftex]{hyperref,graphicx}	
\begin{document}

\title{Comment on \\Y. Couder and E. Fort:\\ {\em Single-Particle Diffraction and Interference at a Macroscopic Scale} \\Phys. Rev. Lett. {\bf 97}, 154101 (2006).}

\author{Anders Andersen$^1$, Jacob Madsen$^1$, Christian Reichelt$^1$, 
        Sonja Rosenlund Ahl$^1$, Benny Lautrup$^2$, Clive Ellegaard$^3$,  
        Mogens T. Levinsen$^3$, and Tomas Bohr$^1$} 
\affiliation{$^1$Department of Physics and Center for Fluid Dynamics, 
                 Technical University of Denmark, DK-2800 Kgs. Lyngby, Denmark \\
$^2$Niels Bohr International Academy, The Niels Bohr Institute \\
$^3$The Niels Bohr Institute, University of Copenhagen, Denmark
}

\vspace{5mm} 

\maketitle
\noindent
In a paper from 2006, Couder and Fort \cite{CouderDoubleSlit} describe
a version of the famous double slit experiment performed with drops
bouncing on a vibrated fluid surface, where interference in the particle
statistics is found even though it is possible to determine
unambiguously which slit the ``walking" drop passes. It is one of the
first papers in an impressive series, showing that such walking
drops closely resemble de Broglie waves and can reproduce typical
quantum phenomena like tunneling and quantized states
\cite{Tunneling,PNAS2010,EddiJFM,Splitting,BushCouder2013,MolacekBushIa,MolacekBushIb,OzaRosalesBush2013,
Oistein,HarrisBushRotate,OzaHarrisBushRotate,CentralForce}. The
double slit experiment is, however, a more stringent test of quantum
mechanics, because it relies upon superposition and phase coherence. In the present
comment we first point out that the experimental data presented in
\cite{CouderDoubleSlit} are not convincing, and secondly we argue 
that it is not possible in general to capture quantum
mechanical results in a system, where the trajectory of the particle is
well-defined. 

In the double slit experiment \cite{CouderDoubleSlit}, 75 drop passages of the slits
are recorded (their Fig. 3). This small number is increased by
symmetrization, which, however, does not improve the
statistics. Submitting the data to a standard $\chi^2$-test, a
fit to a Gaussian distribution is found to be just as good as the fit to the 
Fraunhofer interference pattern
presented in the paper. In addition the blue envelope curve (single
slit result) shown in their Fig. 3 is not backed up by data because
the single slit results presented in the paper (their Fig. 2) are for
slits of different widths than those of their Fig. 3. We have tried to
reproduce their results experimentally with our own double slit set-up, 
but without success.

The walking drops are reminiscent of de Broglie waves. In
his later years de Broglie \cite{deBroglie1987} took his wave idea further
and imagined that particles could be described as moving singularities
in a field, which, in addition to the probabilistic Schr{\"o}dinger
wave function, had a new ``physical" component
excited locally by the particle - just like what happens in the
experiment. We have tried to implement this idea by 
introducing a source term in the standard Schr{\"o}dinger
equation, i.e., $\left(i \, \hbar \, \partial/\partial t - \hat{H}
\right) w({\bf r},t)=J({\bf r}-{\bf R}(t))$ where ${\bf R}(t)$ is the
position of the particle, and the source term $J$ is a
complex function. Due to linearity, it is sufficient to choose the
source term equal to a $\delta$-function, i.e., 
$J({\bf r}-{\bf R}(t)) = \delta({\bf r}-{\bf R}(t))$. In addition, the particle
is guided by the wave according to the standard Madelung-Bohm equation
$\dot {\bf R}(t)= \left(\hbar/m\right) \left.\nabla\Phi({\bf r},t)\right|_{{\bf r} = {\bf R}(t)}$, where
$\Phi$ is the phase of the complex function $w$. In the absence of potential
energy, our theory indeed leads to free ``particles" in the
form of singular wave-fields moving with constant velocity and
resembling the walking drops.

\begin{figure}
\begin{center}
\includegraphics[width=10cm]{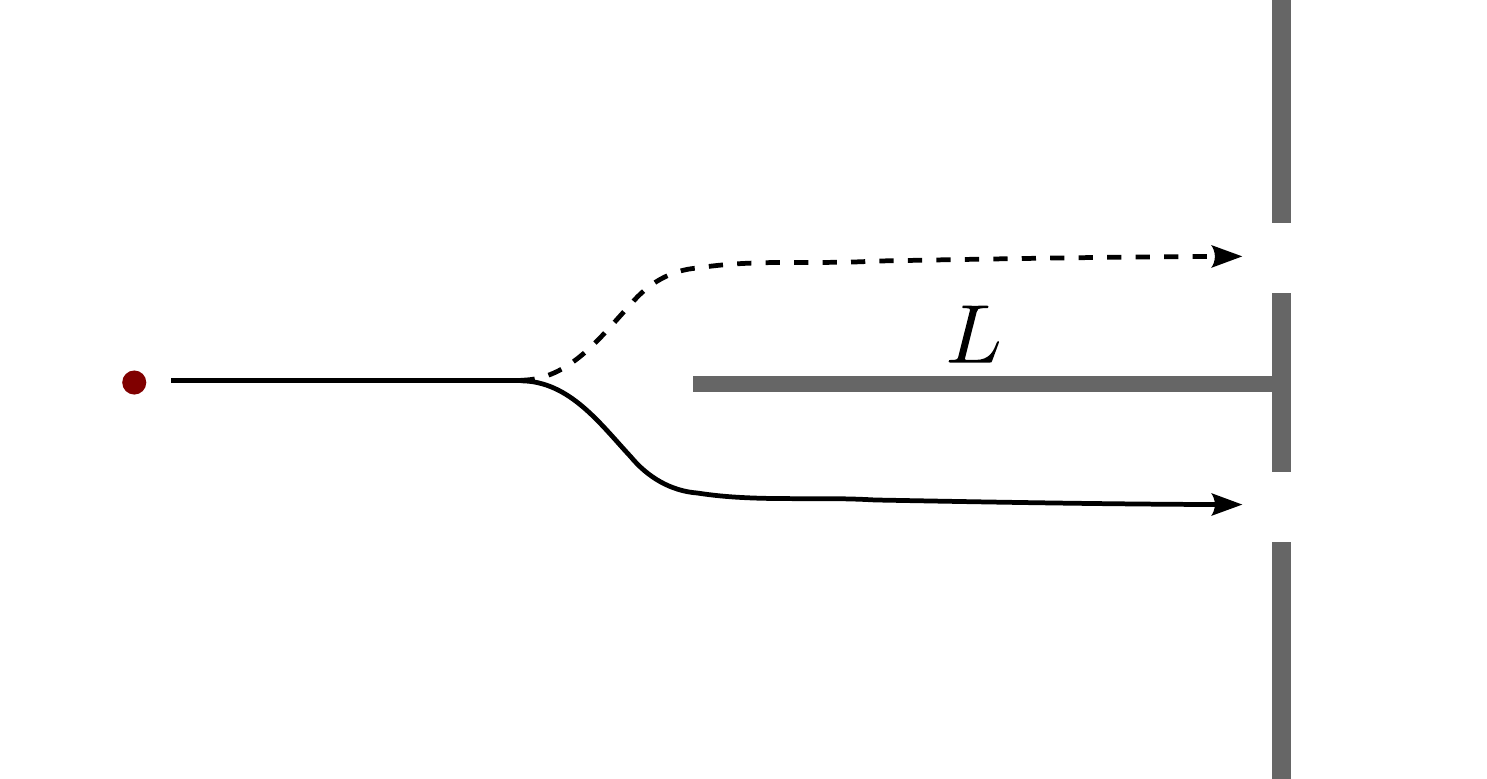}
\caption{\label{splitter}
Double slit experiment with a splitter plate of length $L$. 
A particle will have to move on one or the other side of the plate 
as shown by the two possible trajectories.
}
\end{center}
\end{figure}

We now consider a slightly modified version of the double slit
experiment, where a ``splitter plate" of length $L$ has been inserted
symmetrically in between the two slits (Fig.~\ref{splitter}). 
This change would not significantly alter the quantum mechanical
description or the interference pattern. A wave packet moving towards
the double slit will be diffracted by the edge of the splitter plate and
slowly disperse while moving along it. Thus, with the splitter plate it will be
weaker when it arrives at the double slit, but the two parts of the wave
packet arriving at the slits will still be exactly in phase, thus
giving rise to the same kind of interference pattern as without the
plate.

For our version of de Broglie like quantum mechanics, this would not
be so. The particle reaching the splitter plate would unavoidably have to
proceed along one side or the other (Fig.~\ref{splitter}). The field at the ``chosen'' side
behaves roughly like our stationary solution for a free particle 
while it moves towards the slit. The field on the ``other'' side, however,
has basically no source term, since the source is on the chosen side of
the splitter plate, and the wave-packet on the other side will therefore slowly disperse. 
For sufficiently large $L$, the particle emerging from the
slit will thus only experience an extremely weak influence from the
other slit, which would not be able to deflect it, and as $L \to
\infty$ the classical result is recovered: the superposition of the
probability distributions for each single slit without interference.

\end{document}